\documentclass[a4paper]{jpconf}
\usepackage{graphicx}
\usepackage{iopams}
\usepackage{amsmath}
\usepackage{wasysym}
\usepackage{acronym}

\begin{document}

\title{Bilayer graphene: gap tunability and edge properties}

\author{Eduardo V Castro$^{1}$, N~M~R~Peres$^{2}$, 
J~M~B~Lopes~dos~Santos$^{1}$, F~Guinea$^{3}$ and A~H~Castro~Neto$^{4}$}

\address{$^{1}$ CFP and Departamento de F\'{\i}sica, Faculdade de Ciências
Universidade do Porto, P-4169-007 Porto, Portugal}

\address{$^{2}$ Center of Physics and Departamento de F\'{\i}sica, Universidade
do Minho, P-4710-057 Braga, Portugal}

\address{$^{3}$ Instituto de Ciencia de Materiales de Madrid, CSIC, 
Cantoblanco, E-28049 Madrid, Spain}

\address{$^{4}$ Department of Physics, Boston University, 590 Commonwealth
Avenue, Boston, MA 02215,USA}

\ead{evcastro@fc.up.pt}

\begin{abstract}
Bilayer graphene -- two coupled single graphene layers stacked as in graphite 
-- provides the only known semiconductor with a gap that can be tuned 
externally through electric field effect. Here we use a tight binding approach
to study how the gap changes with the applied electric field. Within a
parallel plate capacitor model and taking into account screening of the 
external field, we describe real back gated and/or chemically doped 
bilayer devices. We show that a gap between zero and midinfrared energies
can be induced and externally tuned in these devices, making bilayer
graphene very appealing from the point of view of applications. However,
applications to nanotechnology require careful treatment of the effect
of sample boundaries. This being particularly true in graphene, where
the presence of edge states at zero energy -- the Fermi level of the 
undoped system -- has been extensively reported. Here we show that also bilayer
graphene supports surface states localized at zigzag edges. The presence of
two layers, however, allows for a new type of edge state which shows an
enhanced penetration into the bulk and gives rise to band crossing
phenomenon inside the gap of the biased bilayer system.
\end{abstract}

%

\section{Introduction}

\label{sec:BilayerIntro}

The recent production of graphene~\cite{NGM+04}, the first truly
one-atom thick material, enabled several unthoughtful
experiments where charge carriers were undoubtedly shown to be massless
with a linear dispersion relation~\cite{NGP+rmp07}.

In addition to single layer graphene (SLG), 
few-layer graphene can also be isolated.
Of particular interest to us is the double layer graphene system,
where two carbon layers are placed on top of each other according
to the usual Bernal $AB$-stacking. The low-energy properties of this
so-called bilayer graphene (BLG)
 are then described by massive Dirac fermions~\cite{NGP+rmp07},
with a quadratic dispersion close to the neutrality point and a Dirac
fermion mass originating from the inter-plane hopping energy $t_{\perp}$.

One of the most remarkable properties of BLG is the ability
to open a gap in the spectrum by electric field effect~--~biased
BLG. This has been realized experimentally,
providing the first semiconductor whose gap can be tuned externally
\cite{OBS+06,OHL+07,CNM+06}.
In the absence of external perpendicular electric field -- unbiased
BLG -- the system is characterized by four bands, two of them
touching each other at zero energy, and giving rise to the massive
Dirac fermions mentioned above, and other two separated by an energy
$\pm t_{\perp}$. Hence, an unbiased BLG is a two-dimensional zero-gap
semiconductor. At the neutrality point
the conductivity shows a minimum of the order of the conductance quantum
\cite{NMcCM+06},
a property shared with SLG~\cite{NGP+rmp07}. This prevents standard
logic applications where the presence of a finite gap producing high
on-off current ratios is of paramount importance. The fact that a
simple perpendicular electric field is enough to open a gap, and even
more remarkable, to control its value, clearly demonstrates the potential
of this system for carbon-based electronics.
A new plateau at zero Hall conductivity \cite{CNM+06} and ferromagnetic
behavior at low densities are some of the gap consequences \cite{CPS+07}.
Moreover, the induced gap 
has been shown to be robust to the presence of disorder \cite{NN06,CPS06}.

Although such a tunable gap by the electric field effect is not possible
in SLG, band gaps can still be engineered by
confining graphene electrons in narrow ribbons \cite{CLR+07,HOZ+07}.
However, the lateral confinement brings about the presence of edges,
which in graphene can have profound consequences on electronics. This
is essentially due to the rather different behavior of the two possible
(perfect) terminations in graphene: \emph{zigzag} and \emph{armchair}.
While zigzag edges support localized states, armchair edges do not
\cite{japonese}. These surface (or edge) states
occur at zero energy, the same as the Fermi level of undoped graphene,
meaning that low energy properties may be substantially altered by
their presence. The self-doping phenomenon \cite{PGN06} and the edge
magnetization with consequent gap opening \cite{japonese} are among
edge states driven effects. In BLG, the question
regarding the presence of edge states is also pertinent. Firstly,
zigzag edges are among the possible terminations in BLG, and
secondly, the presence of edges is unavoidable in tiny devices.

The paper is organized as follows: in Sec.~\ref{sec:BilayerModel}
the lattice structure of BLG and the tight-binding Hamiltonian are
presented; bulk electronic properties are discussed 
in Sec~\ref{sec:BilayerBEP}; the presence of edge states localized
at zigzag edges of BLG is discussed in Sec.~\ref{sec:BilayerSS}; 
Sec.~\ref{sec:conclusions}
contains our conclusions. 

%

\section{Model}

\label{sec:BilayerModel}

\begin{figure}[t]
\begin{centering}
\includegraphics[width=0.9\columnwidth]{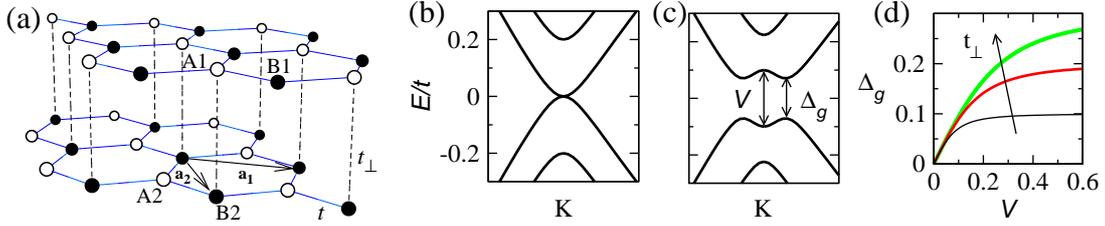}
\par\end{centering}

\caption{\label{cap:bilayer}
(a)~Lattice structure of a graphene bilayer. 
(b)-(c)~Band structure near $K$ (or $K'$) for $V=0$ and $V=t_{\perp}$,
respectively. We used $t_{\perp}=0.2t$.
(d)~Variation of the gap $\Delta_{g}$ with $V$ for 
$t_{\perp}=\{0.1,0.2,0.3\}t$.}

\end{figure}

The lattice structure of BLG is shown in Fig.~\ref{cap:bilayer}(a).
Here we consider only $AB$-Bernal stacking, where the top layer has
its $A$ sublattice on top of sublattice $B$ of the bottom layer.
We use indices~1 and~2 to label the top and bottom layer, respectively.
The basis vectors may be written as
$\mathbf{a}_{1}=a\,\textrm{\^e}_{x}$ and 
$\mathbf{a}_{2}=a(\textrm{\^e}_{x}-\sqrt{3}\,\textrm{\^e}_{y})/2$,
where $a=2.46\,\textrm{\r{A}}$.

In the tight-binding approximation, the in-plane hopping energy, 
$t\approx 3\,\mbox{eV}$,
and the inter-layer hopping energy, $t_{\perp}/t\sim0.1\ll1$ 
\cite{NGP+rmp07}, define the most
relevant energy scales (see Fig.~\ref{cap:bilayer}). The simplest
tight-binging Hamiltonian describing non-interacting $\pi-$electrons
in BLG then reads:
\begin{eqnarray}
H & = & -t\sum_{\mathbf{r},\sigma}\sum_{i=1}^2\big[a_{i,\sigma}^{\dagger}(\mathbf{r})b_{i,\sigma}(\mathbf{r})+a_{i,\sigma}^{\dagger}(\mathbf{r})b_{i,\sigma}(\mathbf{r}-\mathbf{a}_{1})+a_{i,\sigma}^{\dagger}(\mathbf{r})b_{i,\sigma}(\mathbf{r}-\mathbf{a}_{2})+\textrm{h.c.}\big]\nonumber \\
 &  & -t_{\perp}\sum_{\mathbf{r},\sigma}\big[a_{1,\sigma}^{\dagger}(\mathbf{r})b_{2,\sigma}(\mathbf{r})+\textrm{h.c.}\big]+
\frac{V}{2}\sum_{\mathbf{r},\sigma}\big[n_{A1}(\mathbf{r})+n_{B1}(\mathbf{r})-n_{A2}(\mathbf{r})-n_{B2}(\mathbf{r})\big],
\label{eq:Hbilayer}\end{eqnarray}
where $a_{i,\sigma}(\mathbf{r})$ {[}$b_{i,\sigma}(\mathbf{r})$]
is the annihilation operator for electrons at position $\mathbf{r}$
in sublattice $Ai$ ($Bi$), $i=1,2$, and spin $\sigma$, and
 $n_{Ai}(\mathbf{r})$ and $n_{Bi}(\mathbf{r})$ are number operators. 
We are interested in the properties of BLG in the presence of
a perpendicular electric field~--~the biased BLG. The external
perpendicular electric field gives rise to an electrostatic energy
difference between the two layers, which we parametrize by $V$ per
electron. The effect of this energy difference between layers is
 accounted for by the last term in Eq.~(\ref{eq:Hbilayer}).

%

\section{Bulk electronic properties}

\label{sec:BilayerBEP}

In the absence of applied perpendicular electric field 
the system has four bands given by 
\begin{equation}
E_{\mathbf{k}}^{\pm\pm}=\pm\sqrt{\epsilon_{\mathbf{k}}^{2}+t_{\perp}^{2}/4}\pm t_{\perp}/2,
\label{eq:EkBilayer}
\end{equation}
where $\epsilon_{\mathbf{k}}$ is the dispersion of SLG,
$\epsilon_{\mathbf{k}}^2=t\left[3+2\cos(ak_{x})+
4\cos\left(ak_{x}/2\right)\cos\left(ak_{y}\sqrt{3}/2\right)\right]$.
The band structure defined by Eq.~(\ref{eq:EkBilayer})
is shown in Fig.~\ref{cap:bilayer}(b).
In undoped BLG the system has exactly one electron per $\pi$
orbital, that is, it is a half-filled system. In this case the chemical
potential crosses exactly at the $K$ and $K'$ points (the Dirac
points) at the corners of the first Brillouin zone. As can be seen in 
Fig.~\ref{cap:bilayer}(b), at these corners the dispersion
is parabolic, $E(p)\approx\pm v_{\textrm{F}}^{2}p^{2}/t_{\perp}$,
with $p=\hbar q$, where $q$ measures
the distance in momentum space relatively to the Dirac points,
and $v_\textrm{F} = \sqrt3 at\hbar^{-1}/2$ is the SLG
Fermi velocity. Thus,
low-energy quasiparticles in BLG are massive, with a light effective mass 
given by $m^{*}=t_{\perp}/(2v_{\textrm{F}}^{2})\approx0.03m_{\textrm{e}}$,
where $m_{\textrm{e}}$ is the bare electron mass \cite{NGP+rmp07}.
Although this behavior is clearly different from the low-energy massless
Dirac fermions found in SLG, $\epsilon(p)\approx\pm v_{\textrm{F}}p$,
it has been shown that the effective 2-band model describing low-energy
physics in BLG is still a pseudo-spin Hamiltonian \cite{NMcCM+06}.
Consequently, BLG is also far from a standard two-dimensional
electron gas; the observed anomalous quantum Hall effect is an example
of such an unconventional behavior \cite{NMcCM+06}.

Now we address the electronic structure of the biased BLG.
The spectrum of Eq.~(\ref{eq:Hbilayer}) for $V\neq0$ reads: 
\begin{equation}
E_{\mathbf{k}}^{\pm\pm}(V)\!\!=\!\!\pm\sqrt{\epsilon_{\mathbf{k}}^{2}+t_{\perp}^{2}/2+V^{2}/4\pm\sqrt{t_{\perp}^{4}/4+(t_{\perp}^{2}+V^{2})\epsilon_{\mathbf{k}}^{2}}}.
\label{eq:Ekbias}
\end{equation}
The resulting band structure is shown in Fig.~\ref{cap:bilayer}(c).
As can be seen from Eq.~(\ref{eq:Ekbias}), the $V=0$ gapless system
turns into a semiconductor whose gap is controlled by $V$. 
The gap between conduction and valence bands, $\Delta_{g}$, 
is given by
\begin{equation}
\Delta_{g}=\sqrt{t_{\perp}^{2}V^{2}/(t_{\perp}^{2}+V^{2})}.
\label{eq:gapV}\end{equation}
Note that $V$
parametrizes the effect of a perpendicular electric field, and therefore
can be controlled externally. This means, as a consequence of 
Eq.~(\ref{eq:gapV}),
that the biased BLG provides a semiconductor with a gap that
can be tuned externally by electric field effect. From Eq.~(\ref{eq:gapV})
it can be seen that for both $V\ll t_{\perp}$ and $V\gg t$ one finds
$\Delta_{g}\sim V$. However, there is a region for $t_{\perp}\lesssim V\lesssim6t$
where the gap shows a plateau $\Delta_{g}\sim t_{\perp}$, as depicted
in Fig.~\ref{cap:bilayer}(d). The plateau ends when $V\simeq6t$
(not shown).

So far we have considered $V$ as a band parameter that
controls the gap. However, the parameter $V$ can be related with
the perpendicular electric field applied to BLG, avoiding the
introduction of an extra free parameter in the present theory.
If $\mathbf{E}=E\hat{e}_{z}$ is the perpendicular electric
field felt by electrons in BLG, the corresponding electrostatic
energy $U(z)$ for an electron of charge $-e$ is related to the electric
field as $eE=\partial U(z)/\partial z$, and thus $V$ is given by
\begin{equation}
V=U(z_{1})-U(z_{2})=eEd,
\label{eq:VofE}\end{equation}
where $z_{1}$ and $z_{2}$ is the position of layer~1 and~2, respectively,
and $d\equiv z_{1}-z_{2}=3.4\,\mbox{\r{A}}$ is the inter-layer distance.
Given the experimental conditions, the value of $E$ can be calculated
under a few assumptions, as detailed in the following.

%

\subsection{Real biased bilayer devices}

\label{subsec:Eext}

If we assume the electric field $E$ in Eq.~\eqref{eq:VofE} to be
due exclusively to the external electric field applied to BLG,
$E=E_{ext}$, all we need in order to know $V$ is the value of $E_{ext}$:
$V=eE_{ext}d$.
The experimental realization of a biased BLG has been achieved
in epitaxial BLG through chemical doping \cite{OBS+06} and
in back gated exfoliated BLG \cite{CNM+06,OHL+07}. In either
case the value of $E_{ext}$ can be extracted assuming a simple parallel
plate capacitor model.

In the case of exfoliated BLG, devices are prepared by micromechanical
cleavage of graphite on top of an oxidized silicon wafer ($t = 300\,\mbox{nm}$
of $\mbox{SiO}_{2}$).
A back gate voltage $V_{g}$ applied between the sample and the Si
wafer induces charge carriers due to the electric field effect, resulting
in carrier densities $n_{g}=\alpha V_{g}$ relatively to half-filling.
The geometry
of the resulting capacitor determines the coefficient 
$\alpha = \varepsilon_{\textrm{SiO}_{2}}\varepsilon_{0}/(et)
\cong7.2\times10^{10}\,\textrm{cm}^{-2}/\textrm{V}$,
where $\varepsilon_{\textrm{SiO}_{2}}=3.9$ and $\varepsilon_{0}$ are
the permittivities of $\mbox{SiO}_{2}$ and free space, respectively.
In order to control independently the gap value and the Fermi level,
in Ref.~\cite{CNM+06} the devices have been chemically doped by
deposition of ammonia (NH$_{3}$) on top of the upper layer, which
adsorbed on graphene and effectively acted as a top gate providing
a fixed electron density $n_{0}$. Charge conservation
then implies a total density $n$ in BLG given by $n=n_{g}+n_{0}$.
Extending the
parallel plate capacitor model to include the effect of dopants, the
external field $E_{ext}$ is the result of charged surfaces placed
above and below BLG. The accumulation or depletion layer in
the Si wafer contributes with an electric field 
$E_{\textrm{b}}=en_{g}/(2\varepsilon_{r}\varepsilon_{0})$,
while dopants above BLG effectively provide the second charged
surface with electric field 
$E_{\textrm{t}}=-en_{0}/(2\varepsilon_{r}\varepsilon_{0})$,
where $\varepsilon_{r}$ is the bilayer relative dielectric constant.
Adding the two contributions, 
$E_{ext}=E_{\textrm{b}}+E_{\textrm{t}}$,
and making use of the charge conservation relation, we arrive at an
electrostatic energy difference $V$ 
that depends linearly on the BLG density, 

\begin{equation}
V=(n/n_{0}-2)e^{2}n_{0}d/(2\varepsilon_{r}\varepsilon_{0}).
\label{eq:VnUnSgeim}
\end{equation}

In the case of epitaxial BLG, devices are grown on SiC by thermal
decomposition. 
Due to charge transfer from SiC substrate to film, the as-prepared BLG
devices appear electron doped with density $n_{\textrm{a}}$. The
substrate's depletion layer provides the external electric field necessary
to make the system a biased BLG. In Ref.~\cite{OBS+06} the
BLG density $n$ was varied by doping the system with potassium
(K) on top of the upper layer,
which originates an additional charged layer contributing to the external
electric field. Applying the same parallel plate capacitor model as
before, we get an electrostatic energy difference that can be written
as

\begin{equation}
V=(2-n/n_{\textrm{a}})e^{2}n_{\textrm{a}}d/(2\varepsilon_{r}\varepsilon_{0}).
\label{eq:VnUnSohta}
\end{equation}

In the inset of Fig.~\ref{fig:gap}(a)
we compare Eq.~\eqref{eq:VnUnSohta} with experimental results for
$V$ from Ref.~\cite{OBS+06}, obtained by fitting angle resolved
photo-emission spectroscopy (ARPES) measurements.
For this particular biased BLG realization, the as-prepared
carrier density was $n_{\textrm{a}}\approx10\times10^{12}\,\mbox{cm}^{-2}$.
From Eq.~\eqref{eq:VnUnSohta}, this $n_{\textrm{a}}$ value implies
a zero $V$ for the bilayer density 
$n^{\textrm{th}}\approx20\times10^{12}\,\mbox{cm}^{-2}$, 
and therefore zero gap through Eq.~\eqref{eq:gapV}.
Experimentally, a zero gap was found around 
$n^{\textrm{exp}}\approx23\times10^{12}\,\mbox{cm}^{-2}$.
Given the simplicity of the theory, it can be said that $n^{\textrm{th}}$
and $n^{\textrm{exp}}$ are in good agreement. However, the agreement
is only good at $V=0$, since the measured $V$ is not a linear function
of $n$, as Eq.~\eqref{eq:VnUnSohta} implies. This is an indication that
screening due to correlation effects should be taken into account.

%

\subsection{Screening correction and gap behavior}

\label{subsec:Eint}

In deriving Eqs.~\eqref{eq:VnUnSgeim} and~\eqref{eq:VnUnSohta}
we assumed that the electric field $E$ in the BLG region was
exactly the external one, $E_{ext}$. There is, however, an obvious
additional contribution: the external electric field polarizes the
BLG, inducing some charge asymmetry between the two graphene
layers, which in turn give rise to an internal electric field, $E_{int}$,
that screens the external one.

To estimate $E_{int}$ we can again apply a parallel plate capacitor
model. The internal electric field due to the charge asymmetry between
planes may thus be written as
$E_{int}=e\Delta n/(2\varepsilon_{r}\varepsilon_{0})$,
where $-e\Delta n$ is the induced charge imbalance between layers,
which can be estimated through the weight of the wave functions in
each layer, \begin{equation}
\Delta n=n_{1}-n_{2}=\frac{2}{N_{\textrm{c}}A_{\hexagon}}\sum_{j,l=\pm}\sideset{}{'}\sum_{\mathbf{k}}\big(|\varphi_{A1,\mathbf{k}}^{jl}|^{2}+|\varphi_{B1,\mathbf{k}}^{jl}|^{2}-|\varphi_{A2,\mathbf{k}}^{jl}|^{2}-|\varphi_{B2,\mathbf{k}}^{jl}|^{2}\big),\label{eq:Dndef}\end{equation}
where the factor 2 comes from spin degeneracy, $N_{\textrm{c}}$ is
the number of unit cells and $A_{\hexagon}=a^{2}\sqrt{3}/2$ is the
unit cell area, $jl$ is a band label, and the prime sum runs over
all occupied $\mathbf{k}$'s.
Note that in order to calculate $\Delta n$ we must specify
$V$, as it determines the amplitudes on the right hand side
of Eq.~\eqref{eq:Dndef}. On the other hand, $\Delta n$
determines $E_{int}$, and in its turn
$E_{int}$ enters Eq.~\eqref{eq:VofE} through $E$ to give $V$.
Thus, a self-consistent procedure must be followed to calculate the
screened energy difference between layers $V$. In particular, for
the two experimental realizations of biased BLG discussed in
Sec.~\ref{subsec:Eext}, the self-consistent
equation that determines $V$ reads
\begin{equation}
V=[n-2n_0+\Delta n(n,V)]e^{2}d/(2\varepsilon_{r}\varepsilon_{0})
\hspace{0.8cm}\textrm{and}\hspace{0.8cm}
V=[2n_{\textrm{a}}-n+\Delta n(n,V)]e^{2}d/(2\varepsilon_{r}\varepsilon_{0}),
\label{eq:VnS}
\end{equation}
respectively for exfoliated BLG \cite{CNM+06}
and epitaxial BLG \cite{OBS+06} devices. 
It is worth noting that the screening correction expressed 
in Eq.~\eqref{eq:VnS}
leads to a term in the Hamiltonian which is exactly the Hartree correction
due to the charge asymmetry between layers. This approach has been
followed in Refs.~\cite{NNG+06,McC06}, in the continuum approximation,
and has been tested against \emph{ab initio} calculations at half-filling
in Ref.~\cite{MSB+06}.

\begin{figure}[t]
\includegraphics[width=21pc]{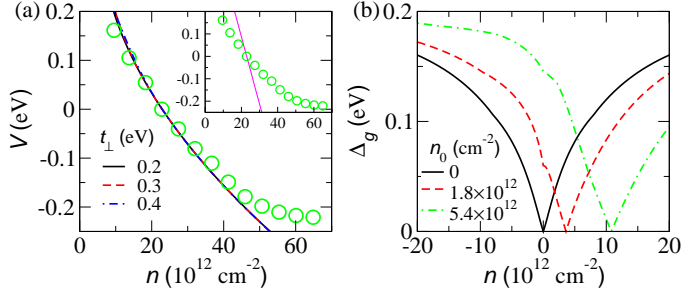}\hspace{2pc}%
\begin{minipage}[b]{14pc}\caption{\label{fig:gap}
(a)~$V$ vs density for the epitaxial
biased BLG device.
Experimental data from Ref.~\cite{OBS+06} is shown as \opencircle.
The line in the inset shows the unscreened result
given by Eq.~\eqref{eq:VnUnSohta}.
(b)~Gap vs density for the exfoliated biased BLG device
 with chemical doping set to $n_0$.}
\end{minipage}
\end{figure}

In Fig.~\ref{fig:gap}(a) we
compare Eq.~\eqref{eq:VnS} (right) with experimental results 
from Ref.~\cite{OBS+06} already 
mentioned in Sec.~\ref{subsec:Eext}. 
 Clearly, the self-consistent
$V$ given by Eq.~\eqref{eq:VnS} for $\varepsilon_{r}=1$ is
a much better approximation than the unscreened result 
of Eq.~\eqref{eq:VnUnSohta}
(see inset). 
In Fig.~\ref{fig:gap}(b) we show the gap $\Delta_{g}$
as a function of carrier density $n$ for the exfoliated biased BLG device,
with realistic
values of chemical doping $n_{0}$ \cite{CNM+06}. The gap is given
by Eq.~\eqref{eq:gapV}, with $t_{\perp}\simeq0.22\,\mbox{eV}$ \cite{CNM+06}
and $V$ obtained by solving self-consistently Eq.~\eqref{eq:VnS} (left)
for $\varepsilon_{r}=1$. 
Note that for $E_{ext}=0$ we always have
$E_{int}=0$ (the charge imbalance must be externally induced), and
therefore we also have $V=0$ and $\Delta_{g}=0$. For this particular
biased BLG device the present model predicts $E_{ext}=0$ for
$n=2n_{0}$, which explains the asymmetry for $\Delta_{g}$ vs $n$
shown in Fig.~\ref{fig:gap}(b). Note also that this device
effectively provides a tunable gap semiconductor, 
as implied by Fig.~\ref{fig:gap}(b):
different gap values are achieved by tuning the gate voltage $V_{g}$,
which controls the carrier density $n$.
Note that the maximum $\Delta_{g}$ occurs
when $V_{g}$ reaches its maximum, which occurs just before the breakdown
of $\mbox{SiO}_{2}$. The breakdown field for $\mbox{SiO}_{2}$ is
$\gtrsim1\mbox{V/nm}$, meaning that $V_{g}$ values as high as $300\,\mbox{V}$
are possible for the exfoliated BLG device.
As shown in Sec.~\ref{subsec:Eext}, $V_{g}\simeq\pm300\,\mbox{V}$
implies $n-n_{0}\simeq\pm22\times10^{12}\,\mbox{cm}^{-2}$, and therefore
Fig.~\ref{fig:gap}(b) nearly spans the interval of possible
densities. 
It is apparent, specially for $n_{0}=5.4\times10^{12}\,\mbox{cm}^{-2}$,
that when the maximum allowed densities are reached the gap seems
to be approaching a saturation limit. This saturation is easily identified
with the plateau shown in Fig.~\ref{cap:bilayer}(d) for $\Delta_{g}$
vs $V$, occurring for $V\gtrsim t_{\perp}$. We may then conclude
that this device enables the entire range of allowed gaps (up to $t_{\perp}$)
to be accessed.

%

\section{Edge states in bilayer graphene}

\label{sec:BilayerSS}

The study of edge states in $AB-$stacked BLG given here is
based on the ribbon geometry with zigzag edges shown 
in Fig.~\ref{cap:ribbon}(a).
Zigzag edges break translational invariance
along their perpendicular direction, enabling us to write an effective
one-dimensional Hamiltonian for a given momentum $ka\in[0,2\pi[$ along
the ribbon. The first nearest-neighbor tight-binding
Hamiltonian [Eq.~\eqref{eq:Hbilayer}] can then be written as%
%
\begin{equation}
H_{k} = -t\sum_{i=1}^2\sum_{n}a_{i;k,n}^{\dagger}(-e^{ika/2}D_{k}b_{i;k,n}+
b_{i;k,n-1}) 
-t_{\perp}\sum_{n}a_{1;k,n}^{\dagger}b_{2;k,n}+
\textrm{h.c.},
\label{eq:Hk}
\end{equation}
where $a_{i;k,n}$ ($b_{i;k,n}$) is the annihilation operator at
momentum $k$ and position $n$ across the ribbon
 in sublattice $Ai$ ($Bi$), $i=1,2$
is the layer index and $D_{k}=-2\cos(ka/2)$.

%

\subsection{Semi-infinite bilayer graphene}

\label{subsec:BilayerSSsemiinf}

Edge states in BLG are investigated by solving the Schr\"odinger equation, 
$H_{k}\left|\psi_{k}\right\rangle =E_{k}\left|\psi_{k}\right\rangle $.
An analytic solution is possible for
the semi-infinite BLG \cite{CPL+07}, the same being true for SLG
\cite{japonese}.
The wavefunction $\left|\psi_{k}\right\rangle $ is written as a linear
combination of the site amplitudes along the edge's perpendicular
direction, $\left|\psi_{k}\right\rangle =
\sum_{n}\sum_{i=1}^2[\alpha_{i}(k,n)\left|a_{i},k,n\right\rangle +
\beta_{i}(k,n)\left|b_{i},k,n\right\rangle ],$
where we have introduced the one-particle states 
$\left|c_{i},k,n\right\rangle =c_{i;k,n}^{\dagger}\left|0\right\rangle $,
with $c_{i}=a_{i},b_{i}$.  Assuming all edge atoms belong
to the $A$ sublattice (without loss of generality),
we require the boundary conditions
$\alpha_{i}(k,n\rightarrow\infty)=\alpha_{i}(k,-1)=
\beta_{i}(k,n\rightarrow\infty)=\beta_{i}(k,-1)=0$, with $i=1,2$,
accounting for the existence of the edge at $n=0$. Within our model,
the Fermi energy of BLG always occurs at zero energy.
Therefore, we expect zero energy edge states to have interesting physical
consequences, and we set $E_{k}=0$. As a result, the two sublattices
become completely decoupled, and only the sublattice to which edge
atoms belong can support edge states. This means that
we always have $\beta_{1}(k,n)=\beta_{2}(k,n)=0$.

\begin{figure}[t]
\begin{centering}
\includegraphics[width=0.98\columnwidth]{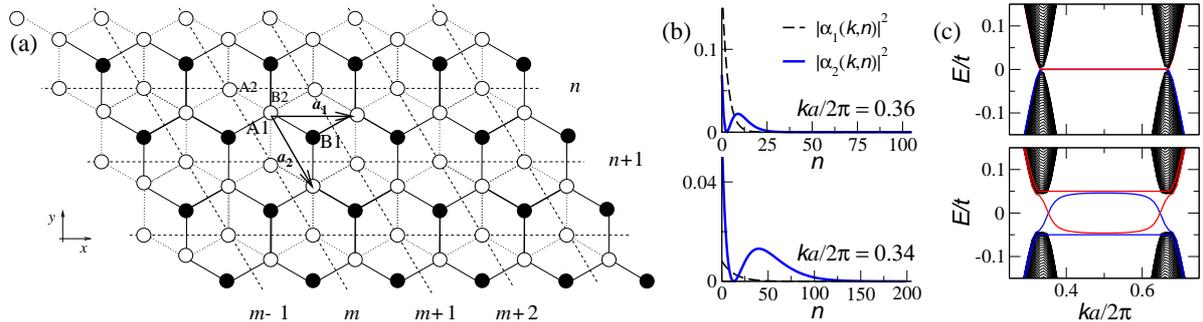}
\par\end{centering}

\caption{\label{cap:ribbon}(a)~Bilayer graphene ribbon with zigzag edges.
(b)~Charge density for bilayer edge states at
$ka/2\pi=0.36$ (top) and $ka/2\pi=0.34$ (bottom).
(c)~Energy spectrum for a bilayer ribbon
with zigzag edges, $N=400$, $t_{\perp}=0.2t$: $V=0$ (top)
and $V=t_{\perp}/2$ (bottom).}

\end{figure}

With $H_k$ as in Eq.~\eqref{eq:Hk}, it is easy to write  
$H_{k}\left|\psi_{k}\right\rangle =0$ as a transfer matrix that relates
$\alpha_{i}(k,n + 1)$ with $\alpha_{i}(k,n)$, for $i=1,2$. Then, applying
the abovementioned boundary conditions we find 
that bilayer graphene supports
two types of zero energy edge states localized at zigzag edges for
$2\pi/3<ka<4\pi/3$: one type restricted to a single layer and coined
\emph{monolayer family}, with amplitudes equivalent to edge states
in SLG,%
%
\begin{equation}
\alpha_1(k,n) = 0 \hspace{0.4cm} \textrm{and} \hspace{0.4cm} 
\alpha_{2}(k,n)=\alpha_{2}(k,0)D_{k}^{n}e^{-i\frac{ka}{2}n};
\label{eq:SSmf}
\end{equation}
and a new type coined \emph{bilayer family}, with finite amplitudes
over the two layers,%
%
\begin{equation}
\alpha_{1}(k,n)  = \alpha_{1}(k,0)D_{k}^{n}e^{-i\frac{ka}{2}n}
\hspace{0.4cm} \textrm{and} \hspace{0.4cm} 
\alpha_{2}(k,n) = -\alpha_{1}(k,0)D_{k}^{n-1}\frac{t_{\perp}}{t}
e^{-i\frac{ka}{2}(n-1)}\Big(n-\frac{D_{k}^{2}}{1-D_{k}^{2}}\Big),
\label{eq:SSbf}
\end{equation}
where the normalization constants are given by 
$|\alpha_{2}(k,0)|^{2}=1-D_{k}^{2}$
and $|\alpha_{1}(k,0)|^{2}=(1-D_{k}^{2})^{3}/[(1-D_{k}^{2})^{2}+
t_{\perp}^{2}/t^{2}]$.
An example of the charge density associated with Eq.~(\ref{eq:SSbf})
is shown in Fig.~\ref{cap:ribbon}(b) for $t_{\perp}=0.2t$, where
the $|\alpha_{1}(k,n)|^{2}$ dependence can also be seen as the solution
given by Eq.~(\ref{eq:SSmf}) for $|\alpha_{2}(k,n)|^{2}$, apart
from a normalization factor.

We note that
the same value of $\lambda$, the penetration depth, is obtained from
Eqs.~(\ref{eq:SSmf}) and~(\ref{eq:SSbf}): $\lambda=-1/\ln|D_{k}|$.
Nevertheless, the solution given by Eq.~(\ref{eq:SSbf}) has a
linear dependence in $n$ which enhances its penetration into the
bulk. We expect these states to contribute more to self doping then
the usual SLG edge states \cite{PGN06}, as the induced Hartree
potential which limits the charge transfer between the bulk and the
edge will be weaker.

%

\subsection{Bilayer zigzag ribbons}

\label{subsec:BilayerSSrib}

So far we studied localized states at the semi-infinite BLG.
Experimentally, however, the relevant situation is a BLG ribbon.
The band structure of a BLG ribbon with zigzag edges
and $N$ unit cells in the perpendicular direction is shown
in the top panel of Fig.~\ref{cap:ribbon}(c), obtained by numerically solving
Eq.~(\ref{eq:Hk}). There are four partly flat bands at $E=0$
for $k$ in the range $2\pi/3\leq ka\leq4\pi/3$, corresponding to
four edge states, two per edge. Strictly speaking, the edge states
given by Eqs.~(\ref{eq:SSmf}) and~(\ref{eq:SSbf}), and the
other two localized at the opposite edge,
 are eigenstates of the semi-infinite system
only. In the ribbon the overlapping of four edge states leads to a
slight dispersion and non-degeneracy. 

The fact that edge states do exist in BLG ribbons imply several
interesting physical properties. First we note that edge states induce
a strong peak in the density of states (DOS) at zero energy, 
due to the flatness of the energy bands. 
This DOS peak can be easily obtained as follows. 
Assuming we are sufficiently far from the Dirac
points, and taking into account that $t_{\perp}/t\ll1$ holds, we
can write for the overlapping between monolayer and bilayer edge state
families $\left|T_{k}^{\textrm{sl}}\right|\approx t\left|D_{k}\right|^{N}$,
while the overlapping between two edge states of the bilayer family
can be written as 
$\left|T_{k}^{\textrm{bl}}\right|\approx t_{\perp}N\left|D_{k}\right|^{N-1}$
(the overlapping between two edge states of the monolayer family being 
identically zero).
The energy dispersion may then be approximated by $E_{k}\approx
\pm\left|D_{k}\right|^{N}[(4t^{2}+
t_{\perp}^{2}N^{2}/D_{k}^{2})^{1/2}\pm t_{\perp}N/\left|D_{k}\right|]$,
which, near $ka\sim\pi$ and assuming $N\gg1$, can be further simplified
as \begin{equation}
E_{k}\sim\begin{cases}
\pm t_{\perp}N\left|ka-\pi\right|^{N}\\
\pm t^{2}\left|ka-\pi\right|^{N}/(t_{\perp}N)\end{cases}.
\label{eq:EkSSap}\end{equation}
Thus the DOS induced by edge states has the form
$\rho(E)\sim E^{-1+\frac{1}{N}}/N$,
which is also found in SLG. Therefore, as predicted for SLG
\cite{WFA+99}, we expect a Curie-like temperature dependence for
the Pauli paramagnetism in BLG due to edge states -- twice
as large due to the presence of twice as many edge states in BLG.%
{} Furthermore, edge magnetism (close to zero temperature) was found 
to be possible in BLG due to edge states \cite{CPSjoam07},
similarly to SLG \cite{japonese}. 

Of particular importance is the effect of a perpendicular
electric field applied to the zigzag bilayer ribbon.
The semi-infinite biased
system has only one edge state given by Eq.~(\ref{eq:SSmf}), as
the edge state having finite amplitudes at both layers 
{[}Eq.~(\ref{eq:SSbf})]
is no longer an eigenstate. In the bottom panel of Fig.~\ref{cap:ribbon}(c) 
we show the band structure of a BLG ribbon for $V=t_\perp / 2$. 
Two partially flat bands for $k$ in the range $2\pi/3\leq ka\leq4\pi/3$
are clearly seen at $E=\pm V/2$. These are bands of edge states localized
at opposite ribbon sides, with finite amplitudes on a single layer
{[}Eq.~(\ref{eq:SSmf})]. Also evident is
the presence of two dispersive bands crossing the gap, which are
reminiscent of the bilayer family of edge states.
This dispersive states appearing inside the gap may contribute to the
finite spectral weight recently observed using ARPES \cite{OBS+06}.

\section{Conclusions}
\label{sec:conclusions}

We have studied the electronic behavior of BLG using the minimal
tight-binding model that describes the system. Particular focus has
been given to the presence of an external electric field perpendicular
to the BLG system -- \emph{biased bilayer} -- which gives rise
to a finite gap in the spectrum, whose size is completely controlled
by the applied voltage. The effect of the perpendicular electric field
has been included through a parallel plate capacitor model, with screening
correction at the Hartree level. 
The biased BLG thus realizes the first tunable gap semiconductor
-- a proof of principle regarding real applications of BLG.
We have also addressed the effect of zigzag edges in BLG. 
We have found that BLG,
as its SLG counterpart and other graphene based materials \cite{CPS08}, 
possesses zero energy surface states
which can be divided into two families, giving rise to twice as many
zero energy bands as in SLG. In the biased case half of the
bands become dispersive inside the gap. 

\section*{References}
\bibliographystyle{apsrev}

\end{document}